\documentclass[fleqn,usenatbib]{mnras}

\usepackage{newtxtext,newtxmath}

\usepackage[T1]{fontenc}

\DeclareRobustCommand{\VAN}[3]{#2}
\let\VANthebibliography\thebibliography
\def\thebibliography{\DeclareRobustCommand{\VAN}[3]{##3}\VANthebibliography}

\newcommand{\teff}{$T_{\mathrm{eff}}$}
\newcommand{\muhz}{$\mu$Hz}
\newcommand{\numax}{$\nu_{\mathrm{max}}$}
\newcommand{\dnu}{$\Delta\nu$}

\newcommand{\kepler}{\textit{Kepler}}

\newcommand{\keplermission}{\textit{Kepler Mission}}

\newcommand{\hst}{\textit{HST}}
\newcommand{\hubble}{\textit{Hubble Space Telescope}}
\newcommand{\stis}{\textit{STIS}}

\newcommand{\msol}{M$_\odot$}

\newcommand{\lsol}{L$_\odot$}
\newcommand{\orcidlink}[1]{\protect\href{https://orcid.org/#1}{\textsuperscript{\protect\includegraphics[width=8pt]{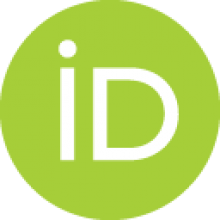}}}}

\usepackage{graphicx}	
\usepackage{amsmath}	

\title[47 Tuc seismology with HST]{Asteroseismology of red giants in the globular cluster 47 Tuc using the HST}

\author[Stello, Bedding \& Gilliland]{
Dennis Stello\orcidlink{0000-0002-4879-3519},$^{1,2}$\thanks{E-mail: d.stello at unsw.edu.au}
Timothy R. Bedding\orcidlink{0000-0001-5222-4661},$^{2}$
and Ronald L. Gilliland\orcidlink{0000-0002-1554-5578}$^{3}$
\\
$^{1}$School of Physics, University of New South Wales, Sydney, NSW 2052, Australia\\
$^{2}$Sydney Institute for Astronomy, School of Physics, University of Sydney, NSW 2006, Australia\\
$^{3}$Space Telescope Science Institute, 3700 San Martin Dr., Baltimore, MD 21218, USA
}

\date{Accepted XXX. Received YYY; in original form ZZZ}

\pubyear{\the\year{}}

\begin{document}
\label{firstpage}
\pagerange{\pageref{firstpage}--\pageref{lastpage}}
\maketitle

\begin{abstract}
Globular clusters provide unique opportunities to study stellar evolution -- as the second brightest cluster, 47 Tuc is a prime target.  Asteroseismology can be used to measure precise masses of stars and has recently been applied to red giants in globular clusters, but so far not for 47 Tuc.  Here, we present a search for solar-like oscillations in red giants of 47 Tuc using 8.3 days of high-cadence Hubble Space Telescope data. We detect oscillations in two out of the five giants falling in the field of view.  One is on the horizontal branch (HB) while the other is on the red giant branch (RGB) at a similar brightness.
From the seismic signal, we measure the stellar masses to be $0.78\pm0.13\,$\msol\ (HB) and $0.94\pm0.15\,$\msol\ (RGB), and hence an inferred integrated mass loss along the upper RGB of $0.16\pm0.20\,$\msol.  A mass uncertainty of less than 0.05\msol\ would be required to obtain a useful estimate of the mass loss, while an uncertainty below 0.01\msol\ would be required to measure the mass difference between the cluster's multiple chemical populations.  The former would be attainable with observations of about 100 times more stars to form ensemble-averaged values, or alternatively a longer campaign observing fewer stars. Detecting mass differences between the chemical sub-populations, could be obtained with a 20-day campaign observing several hundreds of stars.  Our clear detection of oscillations and the prospects presented here warrant dedicated high-cadence campaigns of 47 Tuc, which are possible with NASA's Roman mission and future missions like HAYDN.
\end{abstract}

\begin{keywords}
globular clusters: individual: 47 Tuc 
\end{keywords}




\section{Introduction}
Asteroseismology of red giants arguably began 30 years ago with the detection of oscillations in the globular cluster 47 Tuc. Using the \hubble\ (\hst). \citet{EdmondsGilliland96} reported the discovery of a new class of variable K giants consistent with low over-tone oscillations, with periods in the range $\sim 1$--4 days and amplitudes from 5 to 15 mmag. 
Since that time, there has been a revolution in high-precision space-based photometry \citep{Huber25}, which has driven tremendous growth in asteroseismology of red giants \citep[see reviews by][]{ChaplinMiglio13, GarciaStello15, HekkerChD17, Jackiewicz2021}.  
Based on results from CoRoT, \kepler\ and TESS, it is now clear that the variable K giants detected by \citet{EdmondsGilliland96} in 47~Tuc are undergoing solar-like oscillations. These are acoustic standing waves that are driven stochastically by surface convection, similar to those observed in the Sun.\footnote{In intensity observations, the granulation signal from convection also contributes significantly to the detected variability.}  Solar-like oscillations are seen in essentially all K giants when high-quality data are available \citep[e.g.,][]{Mosser13,Stello14,Yu20}. 

The \citet{EdmondsGilliland96} results were based on only 38.5~hr of \hst\ observations of the cluster core from a 1993 campaign, when the telescope still suffered from spherical aberration.  \hst\ observed 47~Tuc again in 1999\footnote{Program GO-8267}, this time for much longer (8.3\,d) and with new optics that corrected the spherical aberration.  The cluster core was observed with \textit{Wide Field and Planetary Camera 2} (\textit{WFPC2)} to search for transiting exoplanets \citep{Gilliland00} and eclipsing binaries \citep{Albrow01}, but these data were unsuitable for asteroseismology because the red giants were saturated.  However, parallel images were obtained with the Space Telescope Imaging Spectrograph (\stis) for a field offset from the core of 47~Tuc. Here, we analyse these unpublished \stis\ data to search for solar-like oscillations in the red giants, and to inform potential observations of 47~Tuc and other globular clusters with the Roman Space Telescope \citep{Spergel2015, Gould15WFIRST, Huber23, Weiss25} and future missions like HAYDN \citep{Miglio21}.

\section{Observations and analysis}
\subsection{Observations}
\begin{figure}
	\includegraphics[width=\columnwidth]{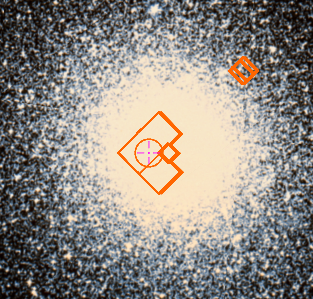}
    \caption{Field of view of 47 Tuc (12x12 arcmin centered on $\mathrm{RA}=00$:24:06.5, $\mathrm{Dec}=-72$:04:38). The \hst\ fields for program GO-8267 are shown in orange. The \stis\ field studied in this paper, centered at $\mathrm{RA}=00$:23:24, $\mathrm{Dec}=-72$:01:24, is in the upper right.} 
\label{fig:fov}
\end{figure}

We analysed \hst\ observations from the \stis\ instrument taken during 1999 July 3--11 (120 orbits), which provide a near-continuous 8.3-day time series. The 50x50 arcsec field was about 4.6 arcmin off the cluster centre (see Figure~\ref{fig:fov}) and included observations with the LONGPASS filter (bandpass about 550--1030nm) and with a CLEAR filter (i.e., no filter; bandpass about 164--1030nm).

For these \stis\ observations, the CCD gain was chosen such that saturation occurred on-chip (as later adopted by the \keplermission; \citealt{Koch10}).  For the \textit{WFPC2} images, on the other hand, counts were clipped by the A/D converter and saturation set in at the bottom of the red giant branch \citep{Albrow01}. Thus, for the brightest stars that could be measured, the \stis\ precision was below 1 ppt per data point, compared to 1--3\% for \textit{WFPC2}. 


\begin{figure}
	\includegraphics[width=\columnwidth]{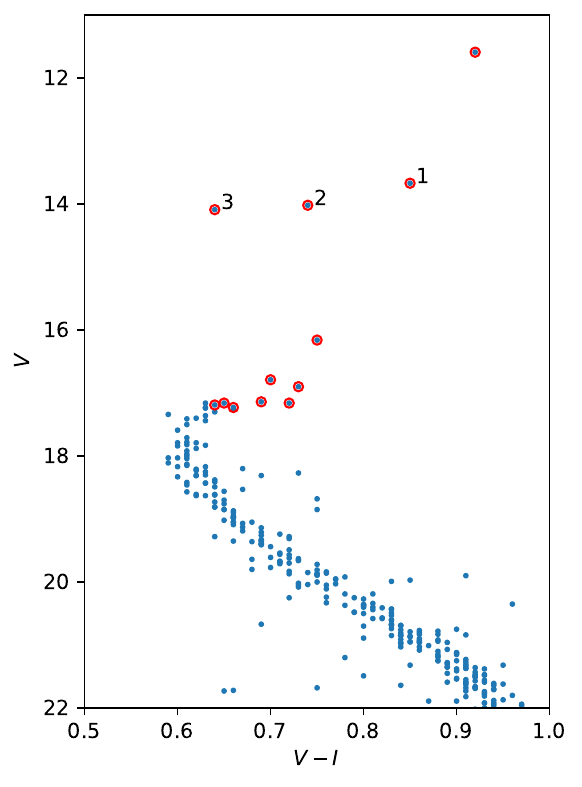}
    \caption{Colour-magnitude diagram of 47 Tuc for the stars with time series data. Stars for which light curves were investigated in detail are shown with red circles.}
\label{fig:cmd-hst}
\end{figure}

The \stis\ photometry was extracted for 1007 stars following the description in \citet{Albrow01} and references therein.  Figure~\ref{fig:cmd-hst} shows the colour-magnitude diagram. The $V$ and $I$ magnitudes were estimated from the \hst/\stis\ data using approximate transformations and were only used to decide which stars to examine in more detail (see next section).
The four brightest stars required special apertures to capture the charge bleeds on the CCD, following \citet{Gilliland94} and similar to the method later adopted for \kepler\ \citep{Gilliland10SC}. We removed correlations of the relative time series photometry with external instrument parameters, including $(x,y)$ positions and a measure of telescope focus.

\subsection{Time series and Fourier analysis}

With a point-to-point photometric precision of about 1~ppt and only 8.3-day time series, we only expect to have a chance of detecting oscillations in stars that have evolved beyond the turnoff \citep{Chaplin11a}.  
We inspected the light curves and power spectra of the stars marked with red circles in Figure~\ref{fig:cmd-hst}.  
The brightest star showed variability that was too slow to characterize its time scale and amplitude, as expected for such a luminous red giant.  The 8 stars at the base of the red giant branch did not show variability, but we did confirm that the CLEAR filter time series had lower scatter than for LONGPASS (by about 20-30\%), reflecting the difference in photon counts.  That leaves the three labelled stars as the best targets for detecting solar-like oscillations.

\begin{figure}
	\includegraphics[width=\columnwidth]{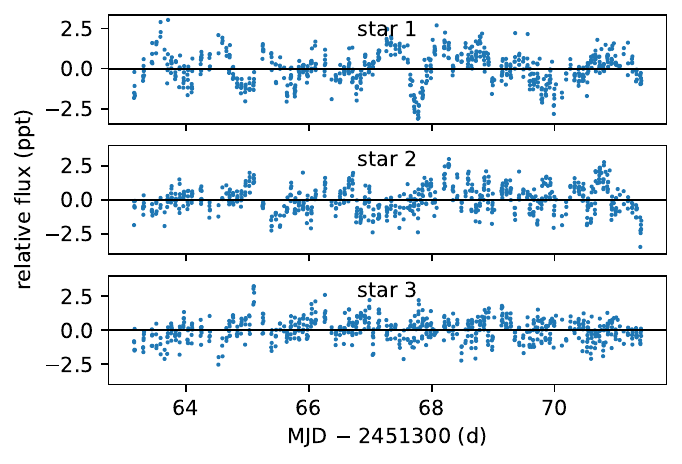}
    \caption{   Light curves from the CLEAR filter for the three stars with a brightness around $V=14$. }
\label{fig:lcs}
\end{figure}

In Figure~\ref{fig:lcs} we show the CLEAR filter light curves of stars 1, 2 and 3, which all demonstrate a point-to-point scatter of around 0.5--0.7 ppt.   
 
\begin{figure}
	\includegraphics[width=\columnwidth]{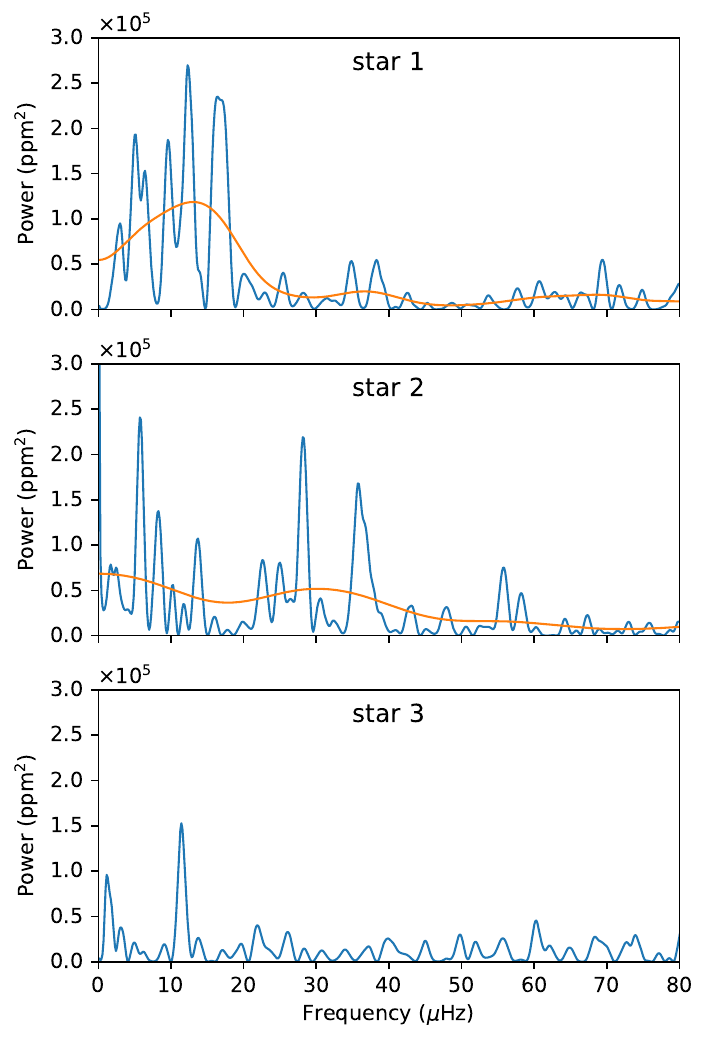}
    \caption{Power spectra of star 1, 2 and 3. The first two spectra smoothed by a Gaussian of FWHM=10\muhz\ are shown in orange.}
\label{fig:power}
\end{figure}
The power spectra are shown in Figure~\ref{fig:power}.
Stars 1 and 2 show variability on timescales of $\sim$20~hr (15~\muhz) and $\sim$8~hr (35~\muhz), respectively, with broad humps in the power spectra that are roughly as expected for stars at these evolutionary stages \citep{Yu18}. Star~2 also shows a rise in power towards low frequencies that is typical of granulation (the oscillation frequency for star~1 is too low for the granulation background to be distinguished with such a short data set).  Meanwhile, Star 3 shows a sawtooth variability with a period of 24~hr (and a single peak in power at 11.57~\muhz), which is presumably a remnant of Earth shine, also seen at the same (or opposite) phase in light curves of other stars we inspected. 
In summary, we find that stars 1 and 2 show oscillations, while star 3 shows only the 1 cycle-per-day peak from Earth shine with no evidence of oscillations or granulation. This is consistent with it being a foreground dwarf, as further discussed in Sec.~\ref{sec:gaia}. 

The next step was to measure \numax\ for stars 1 and~2.
The relatively short time series meant that we could not directly apply the methods developed for \kepler\ data \citep[e.g.][]{Huber09}.
Instead, we used the approach by \citet{Stello17SONG} but with a power-law noise model following \citet{Stello07}, using a Gaussian smoothing width of 2\dnu, where \dnu\ is derived from the \numax--\dnu\ relation by \citet{Stello09a}. This was performed iteratively, where the initial \numax\ was found using a smoothing width of 10\muhz. We note that the results were quite insensitive to this initial choice.
We found \numax\ values of $14.0\pm2.0\,$\muhz\ for star~1 and $30.9\pm4.4\,$\muhz\ for star 2. We also determined the oscillation amplitudes to be $137\pm44\,$ppm and $139\pm56\,$ppm, respectively, using the method described by \citet{Kjeldsen07} as implemented by \citet{Huber09}.  We note that measuring amplitudes from these short time series is challenging and the uncertainties are potentially underestimated. 
The uncertainties in \numax\ and amplitude were determined  following the approach by \citet{Huber09}, which involves perturbing the power spectra with $\chi^2$(2dof) noise \citep{Woodard84}. We did this 100 times and repeated the measurements on each perturbed spectrum, taking the scatter in the measurements across all 100 samples as the uncertainty.  


We note that \citet{Kallinger05} analysed the guide camera data of one star from the same \hst\ campaign that we present here.  They observed a bright giant star with $V=12.235$, hence sitting in the colour-magnitude diagram (Fig.~\ref{fig:cmd-hst}) just below the brighter star in our sample, which we found to show variability too slow to characterise.  \citet{Kallinger05} reported three clear equidistant dominant oscillation frequencies at around 21, 46, and 71\muhz.  However, our detection of a \numax\ of 14\muhz\ in the lower brightness (hence more dense) red giant branch star, 
in agreement with the \numax\ scaling relation (Sect.~\ref{sec:gaia}), 
shows that the \citet{Kallinger05} result is most likely residual variation from Earth shine.  This is further supported by the disagreement between their reported \dnu\ (about 25\muhz) and the dominant oscillation frequencies, which do not follow the expected \numax--\dnu{} relation \citep{Stello09a}.

\subsection{Gaia data and isochrone matching} \label{sec:gaia}
To place our results in context, we examined Gaia DR3 photometry and kinematic information for all stars within 6 arcmin of the cluster centre. This radius includes the observed fields while ensuring that most stars will be cluster members (see Figure~\ref{fig:fov}).  

\begin{figure}
	\includegraphics[width=\columnwidth]{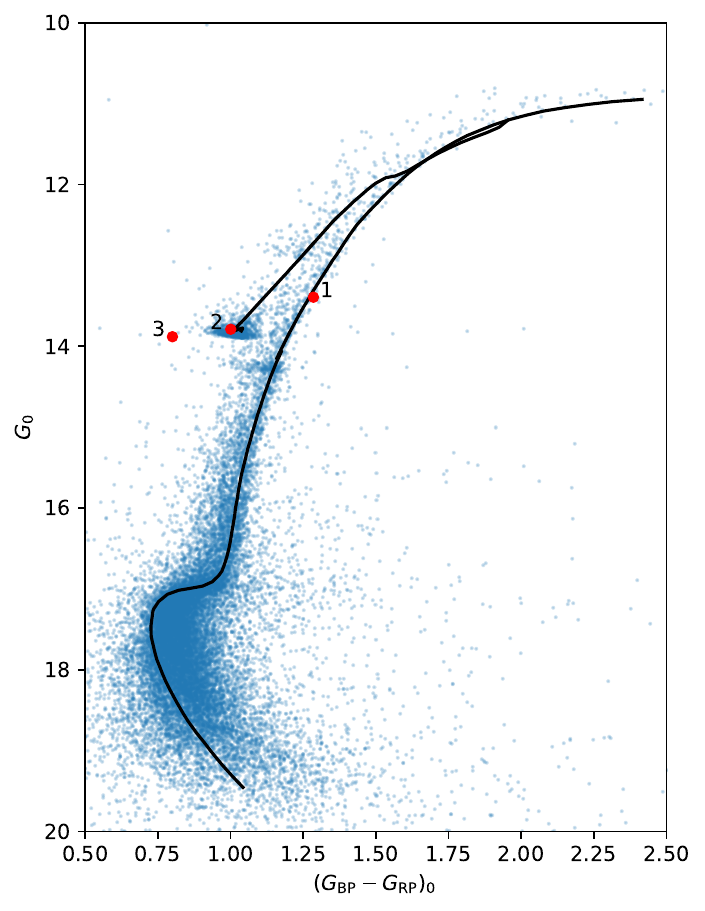}    
    \caption{Colour-magnitude diagram of 47 Tuc based on Gaia DR3 photometry. The PARSEC isochrone (12.75Gyr, $Z=0.0056$) is shifted by a distance modulus of 13.27mag and reddening of 0.04mag. The three stars discussed in the text are highlighted.}
\label{fig:cmd-gaia}
\end{figure}
Figure~\ref{fig:cmd-gaia} shows the Gaia colour-magnitude diagram, with the two oscillating stars (and star 3) highlighted in red.  We also show a PARSEC isochrone\footnote{PARSEC v1.2S +  COLIBRI S37 + S35 + PR16} \citep{Bressan12,Chen14,Chen15,Tang14,Marigo17,Pastorelli19,Pastorelli20} with metallicity 
$Z=0.0056$ 
and an age of 12.75Gyr, including a Reimers mass loss of 0.2 \citep[see][~and reference therein]{Gaiadr2a}. The isochrone was shifted by a distance modulus of 13.27 mag and a reddening of 0.04 mag.

We see that star 1 (Gaia DR3 4689642251832561280) falls on the red giant branch and star 2 (Gaia DR3 4689642148759463040) is on the horizontal branch.  Both have parallaxes and proper motions near the mean of the cluster. This is not the case for star 3 (Gaia DR3 4689642148783280128), whose proper motion is not consistent with cluster membership and whose parallax indicates it is a foreground dwarf. This is in agreement with the lack of granulation and oscillations seen in Fig.~\ref{fig:power}. 

From the location of the two oscillating stars (1 and 2) along the isochrone we can estimate their expected \numax\ using the mass, luminosity, and \teff\ from the isochrone. The widely-used scaling relation \citep{Brown91,KjeldsenBedding95} gives expected \numax\ values of 13.9\muhz\ and 31.0\muhz, respectively, which are in remarkably good agreement with the observations. 

\section{Mass estimations}
With the observed \numax, together with \teff\ and luminosity, one can estimate the stellar mass following the approach by \citet{Stello08} ($M\propto \,$\numax$L/$\teff$^{3.5}$).  
We derive \teff\ using the \textit{colte} method by \citet{Casagrande21} as the weighted average across all colour-\teff\ relations based on both Gaia and 2MASS colour indices. 

To calculate luminosity, we used the following relation
\begin{equation}
    \log(L/\mathrm{L_\odot})=-0.4[K_0+BC-5\log(d)+5-4.75],
\label{eq:lum}
\end{equation}
where we use the cluster distance from \citet{Thompson20} ($4.55\pm0.03\,$ kpc) and reddening from \citet{Brogaard17} ($E(B-V) = 0.03\pm0.01\,$mag)\footnote{Here we used $A_V=3.1E(B-V)$, $A_K=0.114A_V$, and $A_J=0.282A_V$ \citep{Cardelli89}, hence $A_K=0.011$ and $A_J=0.026$ (see also \citealt{Yu26}).}. 
To obtain bolometric corrections, we used \citet{CasagrandeVandenberg18}, adopting their value for the solar absolute bolometric magnitude and [$\alpha/$Fe$]=+0.4$, in agreement with \citet{Brogaard17,Renno20}. Although Eq.~\ref{eq:lum} explicitly lists the $K$ band, we also used $J$ (both from 2MASS), which gave consistent results, to obtain a weighted average luminosity from the two.

Because our measured \numax\ uncertainties are relatively large ($\approx 14\%$), the uncertainties on \teff\ ($\approx 2\%$) and luminosity ($\approx 4\%$) are rather insignificant for the final mass error-budget in this instance.
The resulting seismic masses are $0.94\pm0.15$\msol\ (star 1/red giant branch) and $0.78\pm0.13$\msol\ (star 2/horizontal branch).  We summarise our results and adopted values for the two oscillating giants in Table~\ref{tab:values}.
\begin{table}
	\centering
	\caption{Summary of stellar parameters for the two oscillating giants reported here (see text for details).}
	\label{tab:values}
	\begin{tabular}{lll} 
		\hline
		& Star 1/Red giant branch & Star 2/Horizontal branch \\
		\hline
		Gaia DR3 ID                & 4689642251832561280 & 4689642148759463040 \\
		\numax (\muhz)             & $14.0\pm2.0$        & $30.9\pm4.4$        \\
		Amplitude (ppm)            & $137\pm44$          & $139\pm56$          \\
		\teff (K)                  & $4602\pm108$        & $5258\pm104$        \\
		$L$/\lsol                  & $93.4\pm3.5$        & $55.8\pm2.3$        \\
		$M$/\msol$_\mathrm{,seis}$ & $0.94\pm0.15$       & $0.78\pm0.13$       \\
		\hline
	\end{tabular}
\end{table}

\subsection{Mass loss and multiple stellar populations}
The mass difference between the horizontal branch and the lower part of the red giant branch (near and below the horizontal branch) is a representative measure of the integrated mass loss along the upper red giant branch.  This measure has been obtained using ensemble-averaged asteroseismic masses from \kepler/K2 for several globular clusters (M4 \citet{Howell22,Tailo22}, M80 \citet{Howell24}, M9 and M19 \citet{Howell25}), as well as for the open clusters NGC~6819 \citep{Miglio12,Handberg17}, NGC~6791 \citep{Miglio12}, and M67 \citep{Stello16M67,Reyes25}. 
An intriguing result has recently emerged from this. Both seismic \citep{Howell22,Howell24,Howell25} and non-seismic \citep{Tailo20} results for globular clusters show that mass loss \textit{increases} with increasing metallicity over the range of about $-2.25<\mathrm{[Fe/H]}<-0.50$.  This is in stark contrast to the asteroseismic result from field stars and open clusters \citep{Brogaard24,LiYaguang25} and to the non-seismic results of two globular clusters \citep{Brogaard24}, which show a \textit{decrease} over the range of $-0.9<\mathrm{[Fe/H]}<0.4$.  

Among globular clusters, 47 Tuc has a relatively high metallicity ($\mathrm{[Fe/H]}=-0.72$; \citealt{Harris96}, 2010 edition\footnote{https://physics.mcmaster.ca/\textasciitilde{}harris/Databases.html}), placing it in the contentious metallicity range where the two opposing mass-loss trends overlap but disagree, and where no asteroseismic mass-loss measurement has been determined for any cluster.  
Our results give an estimate of the integrated mass loss of $0.16\pm0.20\,$\msol\footnote{Based on our adopted isochrone, the difference in initial mass between the two stars is only $\simeq0.002\,$\msol.}.  The large uncertainty prevents us from drawing any strong conclusions about mass loss for 47 Tuc.  It is the relatively short time span of the \hst\ observations 
that causes the uncertainty in the mass of each star to be dominated by the \numax\ uncertainty, which scales roughly as the reciprocal of the observing time span \citep{Hekker12}.  However, our clear detections of oscillations are encouraging and warrant dedicated asteroseismic observations aimed at obtaining precise masses of large numbers of red giants in 47~Tuc (see Sec.~\ref{sec:future}). 

In addition to determining mass loss, precise masses from asteroseismology could also determine differences between sub-populations within a single globular cluster.  So far, only tentative results have been obtained \citep{Howell24,Howell25}.  In this regard, 47 Tuc is also an interesting target because it is known to have at least two sub-populations (helium-normal and helium-rich; \citealt{Anderson09,Milone12}), and a small spread in metallicity \citep{Lee22}. 

\section{Requirements for dedicated asteroseismic campaigns}
\label{sec:future}

The difference due to mass loss between the horizontal branch and the lower red giant branch in globular clusters ranges from $\approx 0.1\,$\msol\ to $\approx 0.3\,$\msol, with an expected value of $\approx 0.2$\msol\ for 47 Tuc\ (see references above).  Hence, to draw useful conclusions about the mass lost along the red giant branch, the precision in mass measurements should be less than $\approx 0.05\,$\msol, and in some cases closer to $0.01\,$\msol. 

Determining mass differences between sub-populations would require similar or slightly lower uncertainties than those for mass loss measurements \citep{Tailo20,Howell25}.  For 47 Tuc specifically, \citet{Fu18} fitted isochrones to the two main populations (generations) and found a helium difference of $\Delta Y \simeq 0.02$ (and $\Delta Z \simeq 0.0001$) for their best fitting models ($\eta=0.35$), which equates to a mass difference of $\Delta M \simeq 0.04\,$\msol\ at the main sequence turnoff (see their Table 5).
Also using isochrone fitting, \citet{Tailo20} found a mass difference between the populations on the horizontal branch to be about 0.01--0.02$\,$\msol. This slightly lower value could indicate a difference in the integrated mass loss along the red giant branch for the two populations.
Measuring such small differences would require a precision below $0.01\,$\msol.

One way to reduce the mass uncertainty is to observe for longer. 
With 20 days of observations, our 14\% \numax\ uncertainty, which dominates our current error budget, would be reduced to 5\% (similar to the combined uncertainty from $L$ and \teff).  After 120 days it would reduce to 1\%, making it negligible compared to that of luminosity and \teff.  This would result in a final mass uncertainty of about 4--5\% per star.

Another way to reduce the mass uncertainty is to observe more stars, providing ensemble averages. 
For the 20-day scenario discussed above, observing 3000 giants in 47~Tuc with an equal split between two distinct stellar groups/populations, would result in a mass uncertainty of 0.2\% (or about 0.0015\msol) for each group.  Even observing only 100 stars in each group for 10--15 days would result in an uncertainty of about 1.2\%, allowing strong conclusions on astrophysics related to mass loss and possibly differences among the sub-populations.

\begin{figure}
	\includegraphics[width=\columnwidth]{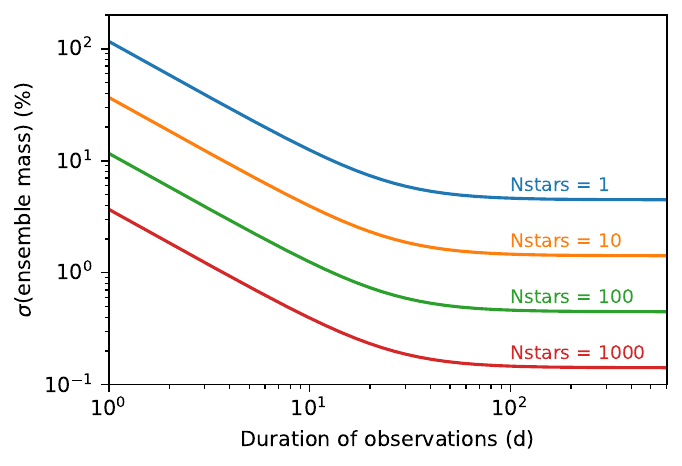}
        \caption{\numax-based mass uncertainty for an ensemble of stars as function of observing duration and size of the ensemble, calibrated to our \hst\ results (Table~\ref{tab:values}, Eq.~\ref{eq:noise}).}
\label{fig:sigmass}
\end{figure}

Observations of 47 Tuc for 10--20 days should be feasible with NASA's Roman mission (currently scheduled for launch in September 2026).  Roman has an aperture similar to that of \hst\ but operates in the near-infrared, and simulations show that the mission should be able to detect oscillations in red giants from saturated photometry \citep{Weiss25}. 
In Fig.~\ref{fig:sigmass} we show how the mass uncertainty of an ensemble depends on the number of stars and the length of the time series.  The plotted relation is
\begin{equation}
\sigma_\mathrm{ensemble~mass}=[( 2^2 + 4^2 + (14\times 8.3/N_\mathrm{days})^2 ) /N_\mathrm{stars}]^{0.5},
\label{eq:noise}
\end{equation}
which is anchored to our \hst/\stis\ results (with uncertainties of 14\% on \numax\ after 8.3 days, 2\% on \teff, and 4\% on luminosity). The relation assumes the \numax\ uncertainty scales inversely with the observation time in agreement with \citet{Hekker12,Stello22,Zinn22}.
In summary, the largest gain in the reduction of the mass uncertainty is within the first 20 days of observations. After this, only a small gain is obtained until about 100 days. Beyond that, masses based on \numax\ no longer improve because they become limited by the \teff\ and luminosity uncertainties (unless one can reduce the uncertainty in \teff\ and luminosity, e.g. using relative measurements). 

In an attempt to obtain more stars from our current \hst\ data, we investigated whether pursuing the \hst\ \textit{WFPC2} observations, obtained during the same campaign as used here, was likely to be fruitful.  Hence, we took the \stis\ CLEAR light curves for the two oscillating giants reported here and added 3\% white noise to simulate the data from the \textit{WFPC2} according to \citet{Albrow01}.  We repeated that to obtain 100 independent light curves for each star, representative of the increase in stars from the \textit{WFPC2} observations, and derived the average power spectrum from those 100 realisations.  Although we knew the frequency range of the oscillations, the spectra showed marginal detections at best, and therefore we did not analyse the \textit{WFPC2} data any further.

Finally, if one were to observe for much longer than 100 days as exemplified above, there is no need to rely on \numax, which has limitations in terms of its precision and accuracy in certain regimes \citep[e.g.][]{Zinn19a,Zinn22,Reyes25}.  Observing for about 1.5 years would enable individual oscillation mode frequencies to be accurately measured, which would open up new avenues in terms of probing stellar structure and evolution.  Not only would the masses of individual stars be more accurate than \numax-based results, but more subtle differences in the oscillation mode frequencies could reveal internal rotation profiles \citep{Beck12} and potentially sharp structural variations in the stellar interiors from mixing processes \citep{Cunha15}.  Such a dedicated program would be ideal for a mission like HAYDN \citep{Miglio21}.

\section*{Acknowledgements}
D.S. and T.R.B. are supported by the Australian Research Council (DP250104267 and FL220100117, respectively).
This research has made use of the SIMBAD database, operated at CDS, Strasbourg, France, and is
based on observations made with the NASA/ESA Hubble Space Telescope, obtained from the MAST data archive at the Space Telescope Science Institute operated by the Association of Universities for Research in Astronomy, Inc., under NASA contract NAS 5-26555.
This work has made use of data from the European Space Agency (ESA) mission Gaia, processed by the Gaia Data Processing and Analysis Consortium (DPAC).

\section*{Data Availability}

The data underlying this article are available on request.



\bibliographystyle{mnras}
\bibliography{bib_complete} 








\bsp	
\label{lastpage}
\end{document}